\documentclass[twocolumn,prc,aps,superscriptaddress,showpacs]{revtex4}
\usepackage{epsfig}
\begin{document}

\title[Inclusive Soft Pion Production]
      {Inclusive Soft Pion Production from 12.3 and 17.5 GeV/c Protons
       on Be, Cu and Au}

\date{\today}

\affiliation{Brookhaven National Laboratory, Upton, NY 11973}
\affiliation{Columbia University, New York, NY 10027}
\affiliation{Florida State University, Tallahassee, FL 32306}
\affiliation{Illinois Institute of Technology, Chicago, IL 60616}
\affiliation{Iowa State University, Ames, IA 50010}
\affiliation{Kent State University, Kent, OH 44242}
\affiliation{Lawrence Berkeley National Laboratory, Berkeley, CA 94720}
\affiliation{Lawrence Livermore National Laboratory, Livermore, CA 94550}
\affiliation{Oak Ridge National Laboratory, Oak Ridge, TN 37831}
\affiliation{State University of New York, Stony Brook, NY 11794}
\affiliation{University of Tennessee, Knoxville, TN 37996}
\affiliation{Yonsei University, Seoul 120-749, Korea}

\author{I.~Chemakin}
\affiliation{Columbia University, New York, NY 10027}
\author{V.~Cianciolo}
\affiliation{Lawrence Livermore National Laboratory, Livermore, CA 94550}
\affiliation{Oak Ridge National Laboratory, Oak Ridge, TN 37831}
\author{B.~A.~Cole}
\affiliation{Columbia University, New York, NY 10027}
\author{R.~C.~Fernow}
\affiliation{Brookhaven National Laboratory, Upton, NY 11973}
\author{A.~D.~Frawley}
\affiliation{Florida State University, Tallahassee, FL 32306}
\author{M.~Gilkes}
\affiliation{State University of New York, Stony Brook, NY 11794}
\author{S.~Gushue}
\affiliation{Brookhaven National Laboratory, Upton, NY 11973}
\author{E.~P.~Hartouni}
\affiliation{Lawrence Livermore National Laboratory, Livermore, CA 94550}
\author{H.~Hiejima}
\affiliation{Columbia University, New York, NY 10027}
\author{M.~Justice}
\affiliation{Kent State University, Kent, OH 44242}
\author{J.~H.~Kang}
\affiliation{Yonsei University, Seoul 120-749, Korea}
\author{H.~G.~Kirk}
\affiliation{Brookhaven National Laboratory, Upton, NY 11973}
\author{N.~Maeda}
\affiliation{Florida State University, Tallahassee, FL 32306}
\author{R.~L.~McGrath}
\affiliation{State University of New York, Stony Brook, NY 11794}
\author{S.~Mioduszewski}
\affiliation{University of Tennessee, Knoxville, TN 37996}
\affiliation{Brookhaven National Laboratory, Upton, NY 11973}
\author{D.~Morrison}
\affiliation{University of Tennessee, Knoxville, TN 37996}
\affiliation{Brookhaven National Laboratory, Upton, NY 11973}
\author{M.~Moulson}
\affiliation{Columbia University, New York, NY 10027}
\author{M.~N.~Namboodiri}
\affiliation{Lawrence Livermore National Laboratory, Livermore, CA 94550}
\author{R.~B.~Palmer}
\affiliation{Brookhaven National Laboratory, Upton, NY 11973}
\author{G.~Rai}
\affiliation{Lawrence Berkeley National Laboratory, Berkeley, CA 94720}
\author{K.~Read}
\affiliation{University of Tennessee, Knoxville, TN 37996}
\author{L.~Remsberg}
\affiliation{Brookhaven National Laboratory, Upton, NY 11973}
\author{M.~Rosati}
\affiliation{Brookhaven National Laboratory, Upton, NY 11973}
\affiliation{Iowa State University, Ames, IA 50010}
\author{Y.~Shin}
\affiliation{Yonsei University, Seoul 120-749, Korea}
\author{R.~A.~Soltz}
\affiliation{Lawrence Livermore National Laboratory, Livermore, CA 94550}
\author{S.~Sorensen}
\affiliation{University of Tennessee, Knoxville, TN 37996}
\author{J.~H.~Thomas}
\affiliation{Lawrence Livermore National Laboratory, Livermore, CA 94550}
\affiliation{Lawrence Berkeley National Laboratory, Berkeley, CA 94720}
\affiliation{Brookhaven National Laboratory, Upton, NY 11973}
\author{Y.~Torun}
\affiliation{State University of New York, Stony Brook, NY 11794}
\affiliation{Brookhaven National Laboratory, Upton, NY 11973}
\affiliation{Illinois Institute of Technology, Chicago, IL 60616}
\author{D.~L.~Winter}
\affiliation{Columbia University, New York, NY 10027}
\author{X.~Yang}
\affiliation{Columbia University, New York, NY 10027}
\author{W.~A.~Zajc}
\affiliation{Columbia University, New York, NY 10027}
\author{Y.~Zhang}
\affiliation{Columbia University, New York, NY 10027}

\begin{abstract}
Differential cross-sections are presented for the inclusive production
of charged pions in the momentum  range 0.1 to 1.2 GeV/c in interactions of
12.3 and 17.5 GeV/c protons with Be, Cu, and Au targets. The measurements
were made by Experiment~910 at the Alternating Gradient Synchrotron in
Brookhaven National Laboratory. The cross-sections are presented as a 
function of pion total momentum and production polar angle $\theta$ with
respect to the beam.
\end{abstract}


\maketitle

\section*{Introduction}

Pions are copiously produced in hadronic interactions at high energies.
However, because of the complexity of soft hadronic interactions,
theoretical descriptions of pion spectra in elementary hadronic
interactions are difficult to obtain. The situation with pion production
in proton-nucleus collisions is even more difficult because of the
contributions from the multiple interaction of the projectile proton
and potential final-state interactions of the outgoing pions with the 
nuclear target. At low momentum these final-state interactions are
expected to be dominated by the low-lying baryon resonances which are 
thought to also play an important role in nucleus-nucleus collisions at 
AGS and lower energies. Precise measurements of pion spectra at low 
momentum may then provide new insight into the role of these resonances and
may help constrain event generators \cite{arc,rqmd} which have been used to
set a baseline for new phenomena in heavy-ion collisions.

Detailed knowledge of pion spectra at low momentum is also of great
practical importance since the production of intense muon beams at
potential new high-energy facilities like muon colliders \cite{mumu} and
neutrino factories \cite{geer,studyII} relies on the decay of pions
produced in a nuclear target. The strong increase in pion yields at very low
momentum resulting from the excitation and decay of baryonic resonances in the
target may provide the necessary yield per beam proton to make such facilities
attractive. Experimental measurement of these spectra is crucial to evaluating
the feasibility of generating such intense beams.

Unfortunately, existing data in the low-momentum region ($<200$ MeV/c)
are limited due to lack of statistics, acceptance, or the particle
identification capabilities of previous experiments
\cite{Abbott,Baker,Dekkers,Lundy,Marmer}. E910 is the first
experiment at the AGS to provide a large statistics data set that
covers a large angular range at low momentum and to provide particle
identification over its entire acceptance. It thus provides an
unprecedented opportunity to fill the gap in our knowledge of pion
production in proton-nucleus interactions.

\begin{figure*}
\includegraphics[width=15cm]{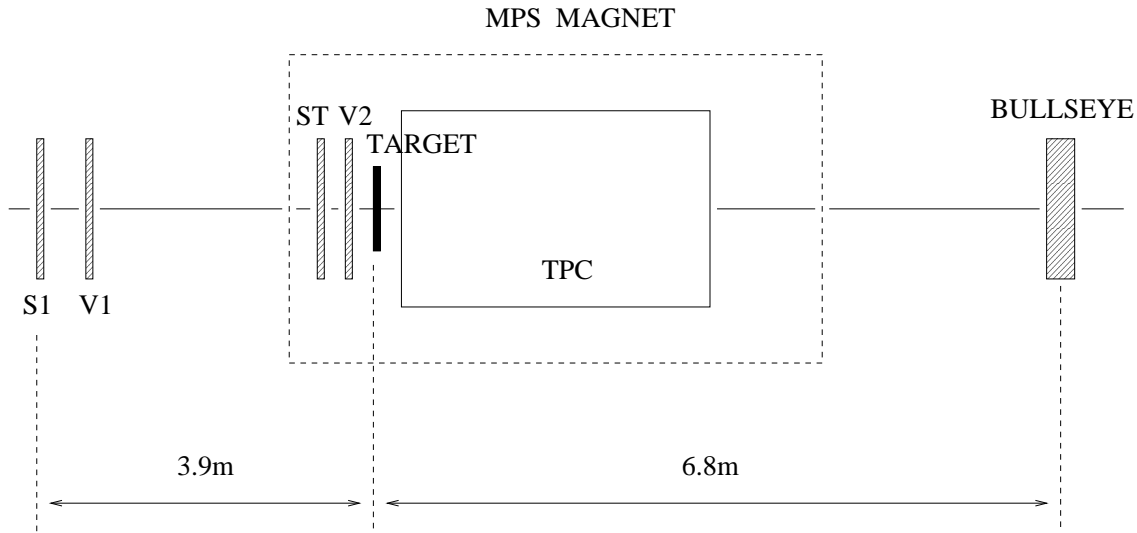}
\caption{E910 detector layout for this measurement (not to scale). Beam comes
         in from the left toward the target located in front of the TPC.}
\label{detconf}
\end{figure*}

\section*{Experiment}

\begin{figure*}
\includegraphics[width=15cm]{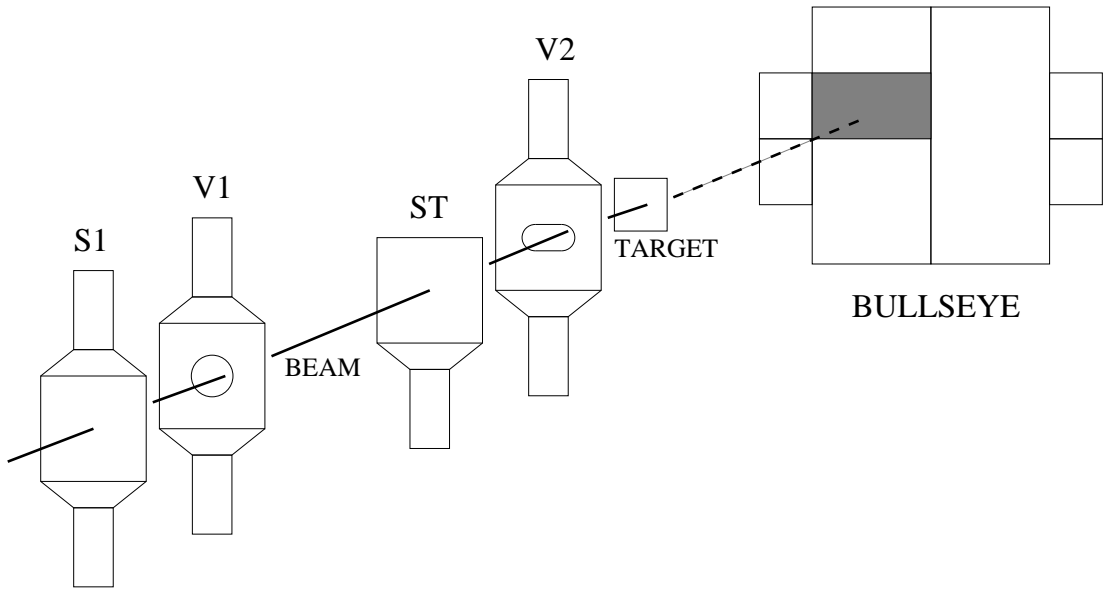}
\caption{Beam trigger for minimum bias production measurement is defined as
         $S1 \cdot \overline{V1} \cdot ST \cdot \overline{V2}$. Shaded
         area on the bullseye defines the beam veto region.}
\label{beamin}
\end{figure*}

Experiment 910 ran for 14 weeks at the MPS facility in the A1 secondary
beamline of the BNL AGS in 1996. Figure~\ref{detconf} shows part of the
detector layout pertinent to the inclusive pion production measurement.
The total momenta and directions of incoming beam particles were reconstructed
using proportional chambers located in the beamline upstream of the target.
Three threshold \v{C}erenkov counters along the beamline were used to reject
particles in the beam other than protons. A set of trigger counters (S1, ST)
and veto counters (V1, V2) located between the proportional chambers and the
target were used to detect and constrain the trajectories of incoming beam
particles. These are shown in Figure~\ref{beamin}. The Au, Cu and Be targets
were rectangular with 5$\times$5, 6.35$\times$6.35 and 7.62$\times$2.54 cm$^2$
cross section and 3.9, 4.1 and 3.4 g/cm$^2$ thickness respectively. The beam
spot on the target was defined by the last veto scintillator which had a
2$\times$1 cm$^2$ slot with semicircular ends.

The EOS time projection chamber (TPC) \cite{TPC} was placed inside the MPS
magnet with its active gas volume about 10 cm downstream of the target
and with its long axis ($z$) approximately aligned with the incoming beam
direction.
It contains 15360 pads of size 0.8$\times$1.2 cm$^2$ arranged in 128 transverse
($x$) rows with 120 pads each to give a 96$\times$153.6 cm$^2$ footprint
for the pad plane. The total vertical drift distance used for recorded ADC
samples was 70 cm. The magnetic field had a strength of 0.5 T and
was oriented along the vertical ($y$) direction. A vertical electric field of
120 V/cm supplied the drift force on electrons from ionization by charged
tracks in the P10 gas at atmospheric pressure. Pad signals were sampled by
12 bit ADCs every 100 ns. Signals from charged tracks were reconstructed as
hits in the TPC centered on padrows in $z$ with the $x$ position given by a
charge weighted mean of the signal from several pads in a row and the $y$
position calculated by the offset from a gamma function fit to the time
dependence. Reconstructed hit positions were corrected for shifts due to
inhomogeneities in the magnetic field before tracking. Simultaneous track
finding and momentum reconstruction with a fixed radius helical fit
\cite{Karimaki} was followed by primary vertex reconstruction \cite{Patrick}
by projecting the beam track forward and TPC tracks back to the target. Track
momenta were then refit using the magnetic field map. The trigger of the
production results in this paper was obtained by requiring the absence of a
beam particle in a downstream ``bullseye'' counter. This counter consisted of
two pairs of scintillator slats, one pair of $14.6 \times 30.5$~cm slats placed
along the vertical and the other $40.6 \times 7.6$~cm pair along the
horizontal. Beam particles consistent with the aperture defined by the veto
counters were entirely located within the intersection of one of the horizontal
and one of the vertical slats as shown in Figure~\ref{beamin}. In addition to
about 4 million minimum bias bullseye triggers, a set of pure beam triggers was
also recorded and used to study potential biases in the bullseye trigger and
to check the normalization of the spectra. A large fraction of bullseye
triggers came from interactions well downstream of the target due to material
in the beam path (including wire chambers and a \v{C}erenkov detector located
between the TPC and bullseye, as well as the TPC gas).

\begin{figure*}
\includegraphics[width=15cm]{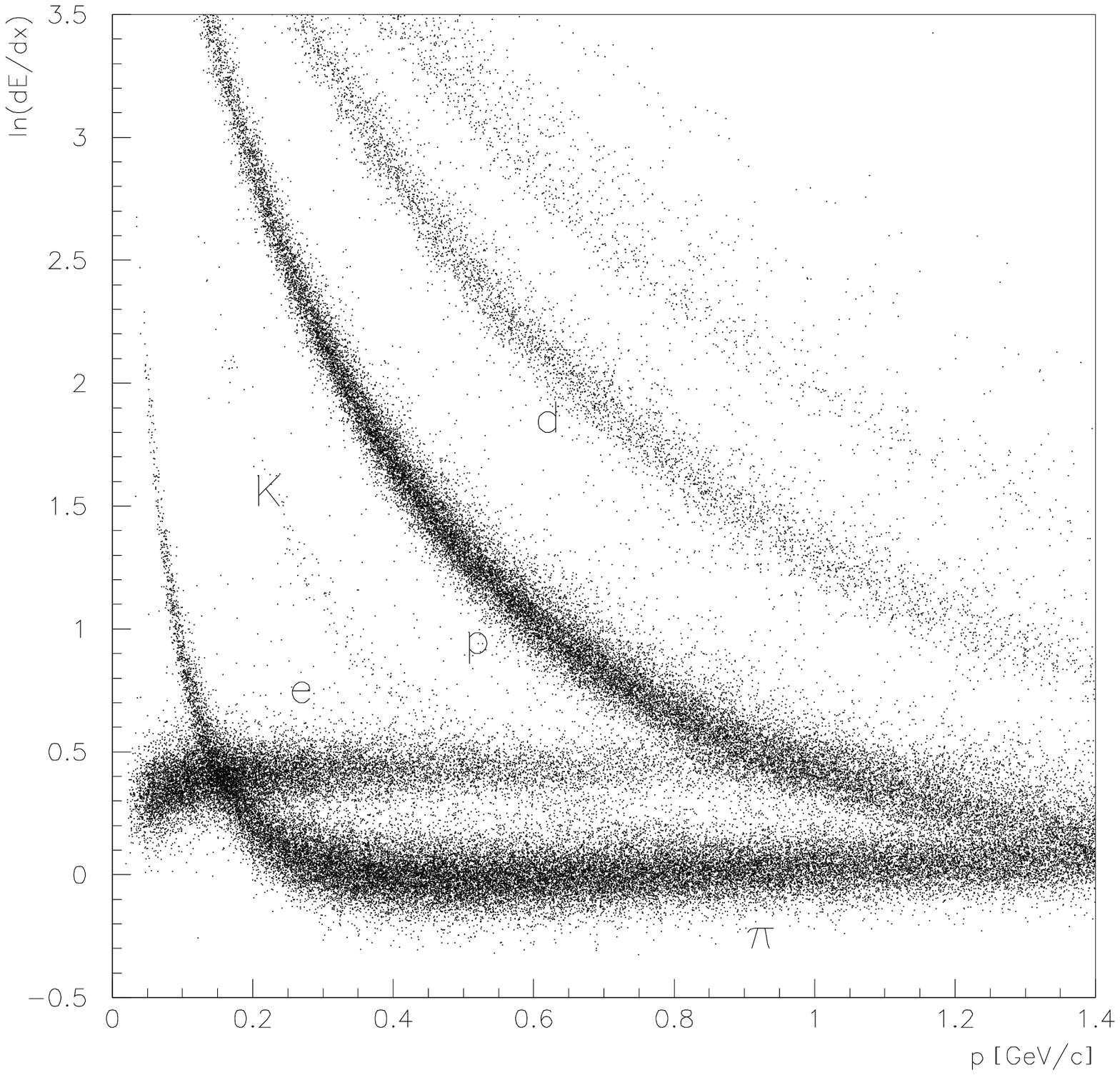}
\caption{$dE/dx$ as a function of momentum showing particles that can be
         identified in different bands.}
\label{dedxvsp}
\end{figure*}

\section*{Data Analysis}

We required a minimum of ten hits in the TPC for charged tracks associated
with the vertex. We also required the reconstructed primary vertex to be
consistent with the target position and the $V2$ veto hole. Interactions were
identified as events with a succesfully reconstructed vertex with at least
two charged tracks in the TPC. Two-prong events consistent with a beam track
and a delta electron from the target were removed from the interaction sample.
After these cuts, there were 16.5-79K events in the bullseye trigger samples
for the six data sets. Table~\ref{intstat} shows the event statistics.
The primary method of particle identification used for slow pions was
ionization energy loss $dE/dx$. Figure~\ref{dedxvsp} shows $dE/dx$ as a
function of momentum below 1.4 GeV/c. Note the overlap of the pion band
with the proton band above 1.2 GeV/c and the overlap of pions in the
100-200 MeV/c range with electrons/positrons resulting primarily from pair
production through photon conversions in the nuclear targets.
The $dE/dx$ for each track was calculated using a 5-65\% truncated mean
after discarding any hits that were shared with other tracks. A residual
shift of the truncated mean with the number of samples was corrected for
using a quadratic fit and the resulting distributions were very nearly
gaussian.

\begin{figure*}
\includegraphics[width=15cm]{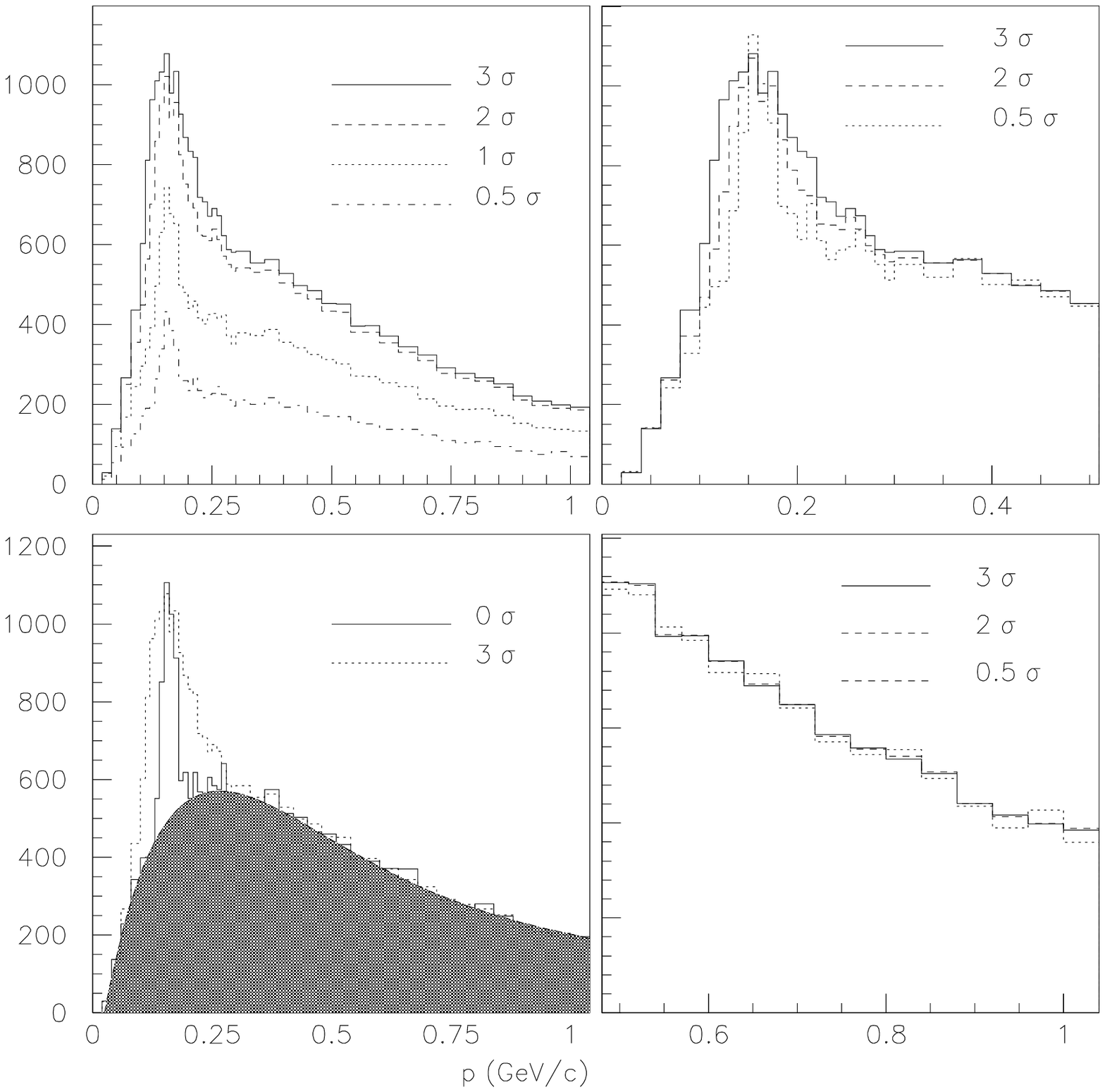}
\caption{Electron background subtraction illustrated in one dimension.
         Top left panel shows momentum distributions in the pion band for
         different widths in $dE/dx$. The top and bottom right panels show
         the low and high momentum half of the histograms obtained by scaling
         these according to the area of a gaussian within a given width. Note
         the shrinking overlap and matching tails. Extrapolation to zero width
         gives the narrowest histogram shown in the bottom left panel
         together with the original $3\sigma$ distribution (dotted line)
         and the fit (solid line) used to remove the background spike on top
         of the smooth pion spectrum (shaded area).}
\label{psigmafit}
\end{figure*}

To resolve the particle identification ambiguity in the low momentum overlap
region, we used the following simple method illustrated in
Figure~\ref{psigmafit}. First we took momentum distributions with progressively
narrower cuts on the $dE/dx$ width around the pion band: $3\sigma$, $2\sigma$,
$1\sigma$, etc. These have the electron background confined to a shrinking
region of momentum. Then we extrapolated the resulting distributions to zero
width after proper scaling. Finally we fit the extrapolated distribution and
subtracted the electron component. In this fit, we used the product of an
exponential and a third order polynomial for the pion distribution and a
gaussian for the electron distribution. The width extrapolation and acceptance
correction were done in two dimensions ($p-\cos \theta$), whereas the
two-component fitting was carried out on one dimensional momentum projections.

\begin{figure*}
\includegraphics[width=16cm]{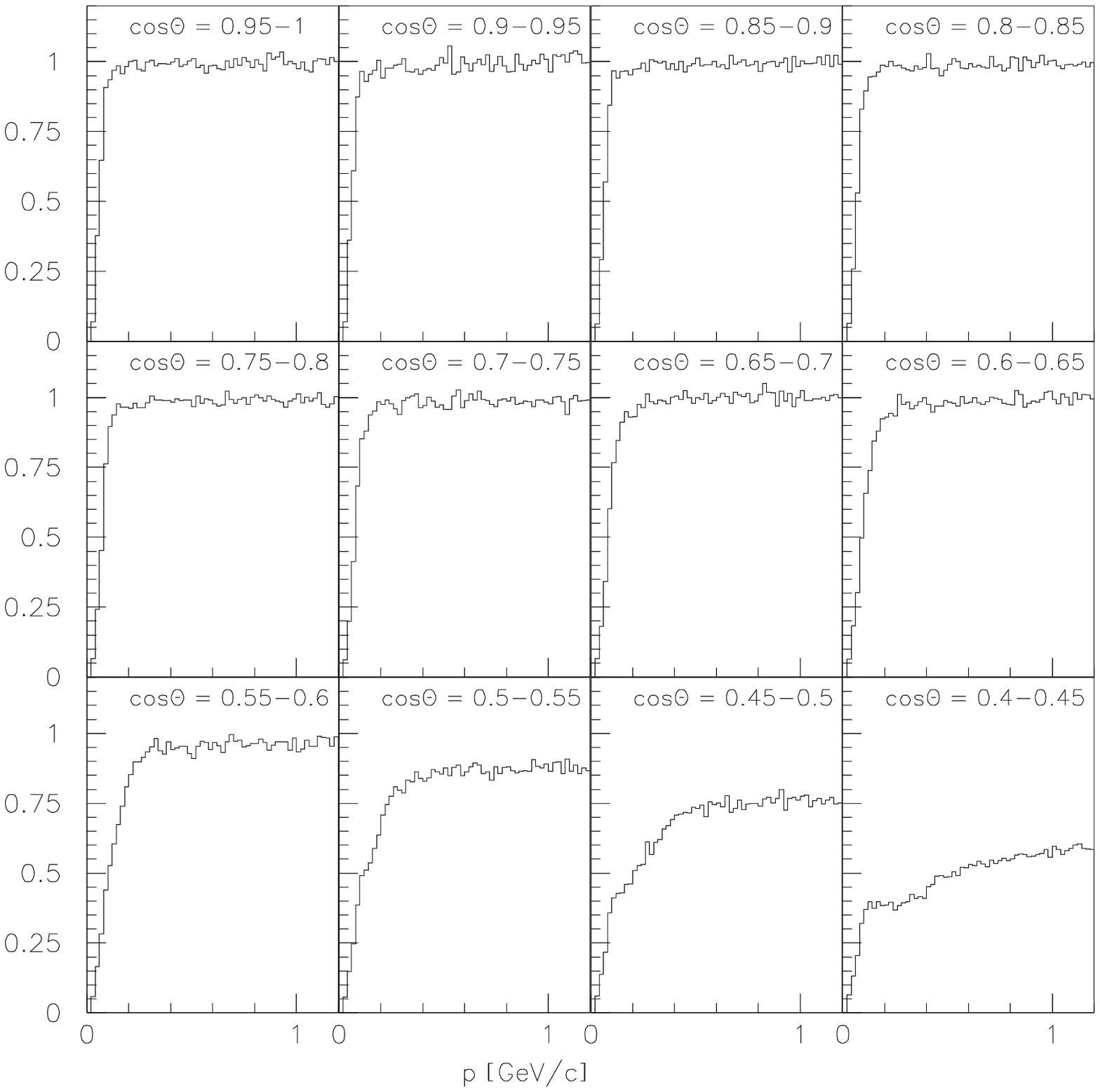}
\caption{TPC acceptance for positive pions coming out of the Au target from
         GEANT simulation as a function of momentum from 20 to 1200 MeV/c in
         0.05 wide $\cos\theta$ bins.}
\label{acceptance}
\end{figure*}

The geometric acceptance of the TPC covers most of the forward hemisphere.
We applied fiducial cuts to the data to exclude regions where the acceptance
is less than 10\%. The full acceptance, shown in Figure~\ref{acceptance} was
calculated by running eight million pion tracks for each target through a
GEANT simulation of our experiment, which included all the relevant processes
(energy loss, multiple scattering, decay) as well as hit and track
reconstruction efficiencies. We also ran a million 17.5 GeV/c p-Au events
generated by RQMD \cite{rqmd} through the simulation to evaluate multitrack
effects.

A lower momentum cutoff of 100 MeV/c was set by requiring the momentum
resolution from multiple scattering and energy loss to be below 5\%. This is
well within the geometric acceptance. The 1.2 GeV/c limit at the other end is
due to the $dE/dx$ overlap between protons and pions. We picked bin sizes in
momentum (100 MeV/c) and $\cos\theta$ (0.1) to keep the statistical uncertainty
in each bin below 20\%. The 100-200 MeV/c bin is further split in two
(100-140 and 140-200 MeV/c) in cases where the electron pion overlap was
restricted to the 140-200 MeV/c region. Our momentum resolution ranges from
2\% at 1.2 GeV/c to 4.5\% at 100 MeV/c. The decay loss in the lowest momentum
bin is less than 4\%. Multitrack effects contribute less than 5\% to the
uncertainty in production results. Also, we estimate the effect of
reinteraction and pion absorption in the target to be less than 5\% each
based on simulations.

\section*{Results}

\begin{figure*}
\includegraphics[width=17cm]{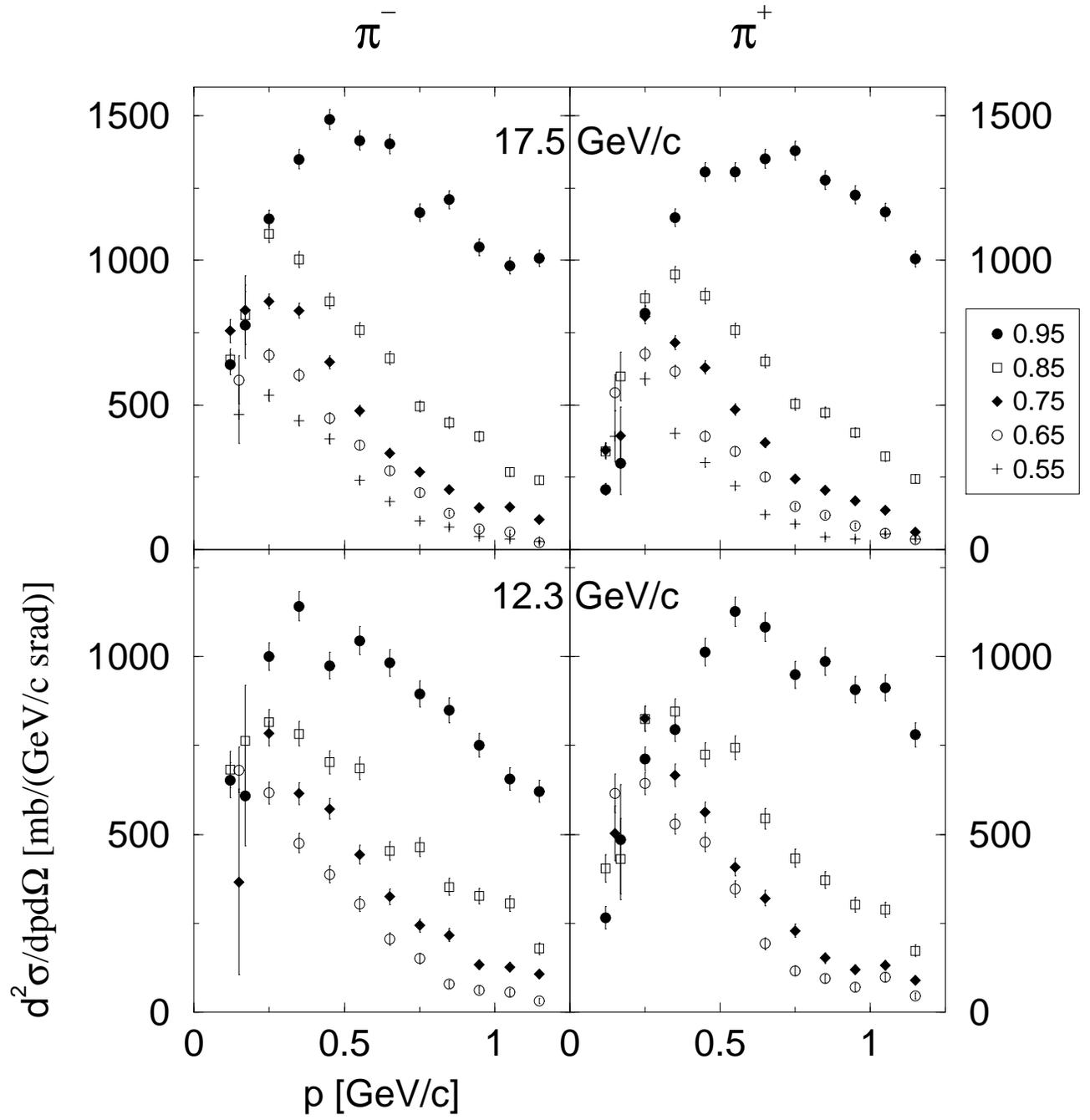}
\caption{Production cross sections for $\pi^{-}$ (left column) and $\pi^{+}$
         (right column) from p-Au at 17.5 (top row) and 12.3 GeV/c (bottom row)
         shown in bins of $\cos\theta$. Numbers in the legend refer to the
         center of each bin.}
\label{aupth}
\end{figure*}

\begin{figure*}
\includegraphics[width=17cm]{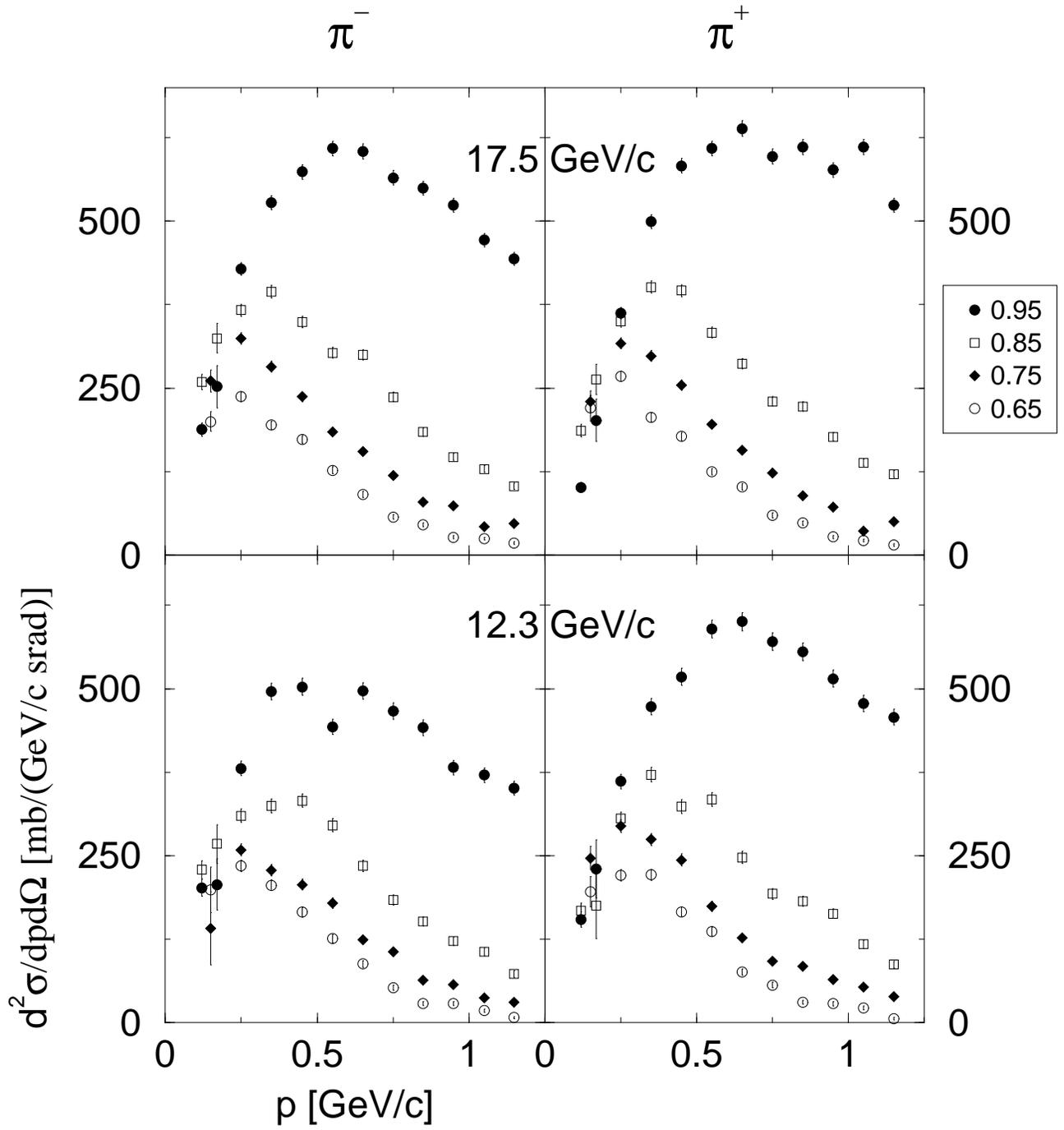}
\caption{Pion production cross sections for p-Cu.}
\label{cupth}
\end{figure*}

\begin{figure*}
\includegraphics[width=17cm]{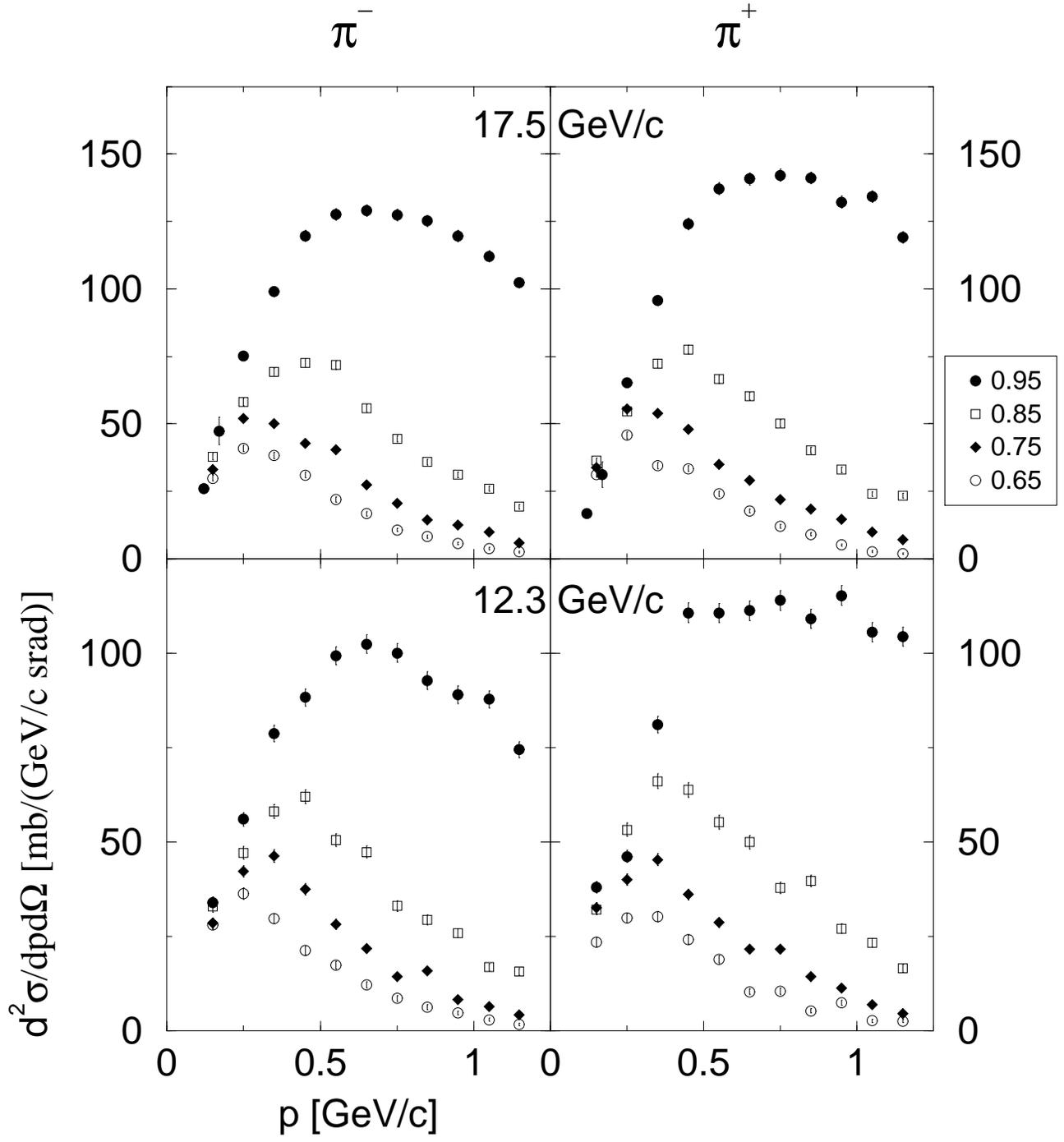}
\caption{Pion production cross sections for p-Be.}
\label{bepth}
\end{figure*}

The differential cross section for pion production for Au, Cu and Be targets
at 12.3 and 17.5 GeV/c are shown in figures \ref{aupth} to \ref{bepth} and
tables \ref{xsau} to \ref{xsbe}. These were calculated in the lab frame
as

\[
\frac{d^2\sigma_{\pi}}{dp\;d\Omega}(p,\cos\theta) = 
\frac{A}{2\pi N_A\rho} \;
\frac{1}{a(p,\cos\theta)} \;
\frac{N_{\pi}(p,\cos\theta)}{N_b\; \Delta p\;\Delta\!\cos\theta}
\]
where $A$ is the atomic weight and $N_A$ is Avogadro's number.
$N_b$ and $\rho$ are the number of incoming beam particles and the area
density of the target. $a(p,\cos\theta)$ stands for the acceptance,
$\Delta p$ and $\Delta\!\cos\theta$ are the bin sizes and 
$N_{\pi}(p,\cos\theta)$ is the number of pions in the bin.
We rescaled the final results by the ratio of measured interaction rates in
the beam and bullseye samples. This correction ranged from 2 to 10\% and is
shown in Table~\ref{intstat}. There is also a 2\% uncertainty in the target
thickness. We estimate the overall normalization uncertainty to be less than
5\%. The errors listed are statistical, but include the uncertainty from
electron background subtraction. After performing the subtraction using a
gaussian fit for the electron contamination, we chose the momentum binning
such that the entire overlap region after width extrapolation was contained in
a single bin. We added the uncertainty in the area of the gaussian fit to the
error bar for the signal in that bin.

\begin{table}
\begin{tabular}{|c|c|c|c|} \hline
Beam        & Target & Number of    & Correction \\
Momentum    &        & Interactions & Factor \\ \hline
            & Au     & 32121 & 1.052 \\ \cline{2-4}
17.5 GeV/c  & Cu     & 54591 & 1.033 \\ \cline{2-4}
            & Be     & 78927 & 1.080 \\ \hline
            & Au     & 16522 & 1.023 \\ \cline{2-4}
12.3 GeV/c  & Cu     & 36891 & 1.096 \\ \cline{2-4}
            & Be     & 41790 & 1.040 \\ \hline
\end{tabular}
\caption{Event statistics and overall normalization factor for
         different data sets}
\label{intstat}
\end{table}

We note that the spectra change from relatively flat momentum distributions
in the forward direction to low momentum peaks at higher angles. Pion
production is mostly forward at higher momenta whereas it is almost isotropic
at lower momenta. This is more pronounced for the heavier Au nucleus than for
the lighter Be nucleus.

An interesting feature of the Au data is the large $\pi^-$ excess over
$\pi^+$ at the lowest momenta, which is larger at the higher beam energy. One
can attribute this in part to the neutron/proton ratio of the Au nucleus. Such
low momentum pions are most likely produced through the decay of resonances
and the intranuclear cascade could favor negative pions over positives. The
effect is diminished for Be and Cu which have about the same number of neutrons
as protons.

Looking at the 17.5 GeV/c p-Au distributions in more detail, for $\pi^-$, the
peak in production shifts from about 450~MeV/c momentum in the most forward
bin to about 250~MeV/c in all larger angle bins. $\pi^+$ distributions are
peaked at higher momenta, around 750~MeV/c in the most forward bin and
350~MeV/c in the second forward bin, while larger angle bins look similar.
$\pi^+$ and $\pi^-$ yields are very close above 500 MeV/c in all except the
most forward bin. The $\pi^-$ excess over $\pi^+$ is most pronounced in the
three most forward $\cos\theta$ bins where the $\pi^-$ to $\pi^+$ ratio in the
100-140 MeV/c momentum bin is 2-3. At larger angles, the $\pi^-$ and $\pi^+$
yields are close. The large $\pi^-$ to $\pi^+$ yield ratio at low momenta and
forward angles warrants further study. There are no significant systematic
effects in our tracking to bias the reconstruction against one sign. Our
acceptance is also insensitive to the sign of track curvature within the
region of momentum space considered in our reported results. Neutral decay
contribution, on the other hand, may be appreciable and we can not rule out the
possibility that some of the asymmetry between $\pi^+$ and $\pi^-$ at low
momentum is due to the decays of slow $\Lambda$'s. We have used the results of
our 17.5~GeV/c p+Au RQMD simulation to evaluate this effect within the context
of that model. Although the contribution of $\Lambda$ decays to the total
$\pi^-$ yield below 1.2~GeV/c is small (about 3\%), it is up to 8\% at
100-300 MeV/c in the most forward $\cos\theta$ bin and lower in the larger
angle bins. The $K_s$ decay contribution to $\pi^+$ and $\pi^-$ yields in this
sample is no more than 2\% of the total in any bin.

\begin{acknowledgments}
We thank Dr. R.~W.~Hackenburg and the MPS staff, J.~Scaduto, and Dr. G.~Bunce
for their help in setting up and running the experiment. We also thank
T. Schlagel and A. Stange for help with computing resources at BNL for data
analysis.
This work was supported by the U.S. Department of Energy under contracts
with BNL (DE-AC02-98CH10886), Columbia (DE-FG02-86ER40281),
ISU (DOE-FG02-92ER4069), KSU (DE-FG02-89ER40531), LBNL (DE-AC03-76F00098),
LLNL (W-7405-ENG-48), ORNL (DE-AC05-96OR22464), and UT (DE-FG02-96ER40982),
and the National Science Foundation under contract with Florida State
University (PHY-9523974). It was also supported in part by grants from
the Illinois Board of Higher Education and the Illinois Department of Commerce
and Community Affairs.
\end{acknowledgments}

\squeezetable
\begin{table*}
\begin{tabular}{ccrrrrrrrr}
 & & \multicolumn{4}{c}{17.5 GeV/c} & \multicolumn{4}{c}{12.3 GeV/c} \\
$\cos\theta$ & $p$
& $d^2\sigma/dpd\Omega$ & error(stat) & $d^2\sigma/dpd\Omega$ & error(stat)
& $d^2\sigma/dpd\Omega$ & error(stat) & $d^2\sigma/dpd\Omega$ & error(stat)\\
& [GeV/c] & $\pi^+$ & \multicolumn{2}{c}{[mb/(GeV/c)]} & $\pi^-$
          & $\pi^+$ & \multicolumn{2}{c}{[mb/(GeV/c)]} & $\pi^-$ \\ \hline
0.95 & 0.12 & 208.3 &  20.5 & 640.4 &  35.9 & 265.8 &  31.5 & 652.2 &  49.3 \\
     & 0.17 & 297.8 & 106.7 & 776.6 & 116.0 & 486.2 & 153.0 & 609.1 & 142.0 \\
     & 0.25 & 817.0 &  25.6 & 1143.0 & 30.3 & 712.6 &  32.6 & 1000.1 & 38.6 \\
     & 0.35 & 1146.9 & 30.4 & 1349.5 & 32.9 &  794.6 & 34.4 & 1140.5 & 41.3 \\
     & 0.45 & 1305.6 & 32.4 & 1487.8 & 34.6 & 1011.7 & 38.9 & 974.1  & 38.1 \\
     & 0.55 & 1304.9 & 32.4 & 1414.5 & 33.7 & 1125.7 & 41.0 & 1043.8 & 39.5 \\
     & 0.65 & 1350.5 & 32.9 & 1402.3 & 33.6 & 1082.4 & 40.2 &  981.5 & 38.3 \\
     & 0.75 & 1378.8 & 33.3 & 1165.3 & 30.6 &  948.0 & 37.6 &  894.9 & 36.5 \\
     & 0.85 & 1277.7 & 32.0 & 1209.3 & 31.2 &  985.8 & 38.4 &  849.0 & 35.6 \\
     & 0.95 & 1225.7 & 31.4 & 1045.3 & 29.0 &  906.8 & 36.8 &  749.9 & 33.5 \\
     & 1.05 & 1167.6 & 30.6 &  980.9 & 28.1 &  912.6 & 36.9 &  655.1 & 31.3 \\
     & 1.15 & 1004.6 & 28.4 & 1007.5 & 28.5 &  779.5 & 34.1 &  621.1 & 30.5 \\
\hline
0.85 & 0.12 & 339.9 & 26.1 &  656.5 &  36.3 & 404.0 &  38.8 & 682.0 &  50.4 \\
     & 0.17 & 599.1 & 84.5 &  811.4 & 102.6 & 430.9 & 114.2 & 763.2 & 155.7 \\
     & 0.25 & 867.8 & 26.4 & 1091.3 &  29.6 & 824.4 &  35.1 & 815.3 &  34.9 \\
     & 0.35 & 951.5 & 27.7 & 1003.6 &  28.4 & 845.7 &  35.5 & 782.7 &  34.2 \\
     & 0.45 & 876.5 & 26.5 &  859.1 &  26.3 & 724.1 &  32.9 & 702.4 &  32.4 \\
     & 0.55 & 758.1 & 24.7 &  759.3 &  24.7 & 743.8 &  33.3 & 685.4 &  32.0 \\
     & 0.65 & 651.6 & 22.9 &  661.8 &  23.1 & 544.2 &  28.5 & 454.1 &  26.0 \\
     & 0.75 & 503.0 & 20.1 &  494.7 &  19.9 & 433.4 &  25.4 & 465.0 &  26.3 \\
     & 0.85 & 472.4 & 19.5 &  438.6 &  18.8 & 371.5 &  23.5 & 352.8 &  22.9 \\
     & 0.95 & 404.5 & 18.0 &  390.5 &  17.7 & 302.2 &  21.2 & 327.1 &  22.1 \\
     & 1.05 & 321.5 & 16.1 &  268.4 &  14.7 & 288.4 &  20.7 & 305.6 &  21.4 \\
     & 1.15 & 245.1 & 14.0 &  240.4 &  13.9 & 173.5 &  16.1 & 179.3 &  16.4 \\
\hline
\end{tabular}
\caption{Production cross sections for p-Au. The bin sizes are 40 and 60 MeV/c
         for the two lowest momentum bins, 100 MeV/c for the rest.}
\label{xsau}
\end{table*}

\begin{table*}
\begin{tabular}{ccrrrrrrrr}
 & & \multicolumn{4}{c}{17.5 GeV/c} & \multicolumn{4}{c}{12.3 GeV/c} \\
$\cos\theta$ & $p$
& $d^2\sigma/dpd\Omega$ & error(stat) & $d^2\sigma/dpd\Omega$ & error(stat)
& $d^2\sigma/dpd\Omega$ & error(stat) & $d^2\sigma/dpd\Omega$ & error(stat)\\
& [GeV/c] & $\pi^+$ & \multicolumn{2}{c}{[mb/(GeV/c)]} & $\pi^-$
          & $\pi^+$ & \multicolumn{2}{c}{[mb/(GeV/c)]} & $\pi^-$ \\ \hline
0.75 & 0.12 & 344.3 &  26.3 & 755.4 &  39.0 &       &      &       &       \\
     &      &       &       &       &       & 502.4 & 77.4 & 366.7 & 260.8 \\
     & 0.17 & 392.7 & 101.1 & 828.1 & 119.4 &       &      &       &       \\
     & 0.25 & 806.2 &  25.5 & 857.3 &  26.2 & 826.4 & 35.1 & 783.6 &  34.2 \\
     & 0.35 & 715.5 &  24.0 & 826.4 &  25.8 & 667.0 & 31.6 & 614.7 &  30.3 \\
     & 0.45 & 629.9 &  22.5 & 648.1 &  22.8 & 562.4 & 29.0 & 571.7 &  29.2 \\
     & 0.55 & 483.5 &  19.7 & 479.3 &  19.6 & 407.8 & 24.7 & 443.4 &  25.7 \\
     & 0.65 & 369.9 &  17.2 & 333.4 &  16.4 & 320.4 & 21.9 & 324.9 &  22.0 \\
     & 0.75 & 244.3 &  14.0 & 268.2 &  14.7 & 229.7 & 18.5 & 244.0 &  19.1 \\
     & 0.85 & 204.5 &  12.8 & 207.3 &  12.9 & 153.0 & 15.4 & 217.5 &  18.0 \\
     & 0.95 & 168.5 &  11.6 & 145.2 &  10.8 & 120.9 & 13.4 & 133.4 &  14.1 \\
     & 1.05 & 135.6 &  10.4 & 146.6 &  10.9 & 132.2 & 14.0 & 128.0 &  13.8 \\
     & 1.15 &  59.8 &   7.8 & 104.5 &   9.2 &  90.7 & 12.6 & 108.0 &  12.7 \\
\hline
0.65 & 0.15 & 541.6 & 62.5 & 586.7 & 83.5 & 615.5 & 54.5 & 680.7 & 64.3 \\
     & 0.25 & 676.2 & 23.3 & 671.1 & 23.2 & 643.3 & 31.0 & 616.5 & 30.3 \\
     & 0.35 & 615.3 & 22.2 & 602.6 & 22.0 & 530.0 & 28.1 & 475.5 & 26.6 \\
     & 0.45 & 392.1 & 17.8 & 454.0 & 19.1 & 478.0 & 26.7 & 387.4 & 24.0 \\
     & 0.55 & 338.4 & 16.5 & 361.0 & 17.0 & 346.6 & 22.7 & 305.0 & 21.3 \\
     & 0.65 & 251.1 & 14.2 & 273.4 & 14.8 & 193.2 & 17.0 & 207.0 & 17.6 \\
     & 0.75 & 149.8 & 11.0 & 196.2 & 12.6 & 117.3 & 13.3 & 151.0 & 15.0 \\
     & 0.85 & 118.8 &  9.8 & 126.3 & 10.1 &  94.9 & 11.9 &  79.7 & 11.4 \\
     & 0.95 &  81.9 &  8.3 &  71.6 &  7.6 &  71.2 & 11.0 &  62.7 &  9.9 \\
     & 1.05 &  55.9 &  6.9 &  59.8 &  6.9 &  99.0 & 12.2 &  56.1 &  9.2 \\
     & 1.15 &  35.4 &  6.0 &  24.5 &  5.0 &  46.9 &  9.0 &  32.3 &  6.9 \\
\hline
0.55 & 0.15 & 391.1 & 88.0 & 467.2 & 100.5 &      &      &       &      \\
     & 0.25 & 589.7 & 21.8 & 533.5 &  20.7 &      &      &       &      \\
     & 0.35 & 402.2 & 18.0 & 445.9 &  18.9 &      &      &       &      \\
     & 0.45 & 300.1 & 15.5 & 382.9 &  17.5 &      &      &       &      \\
     & 0.55 & 219.9 & 13.3 & 239.0 &  13.9 &      &      &       &      \\
     & 0.65 & 122.0 &  9.9 & 165.5 &  11.5 &      &      &       &      \\
     & 0.75 &  88.9 &  8.7 &  99.5 &   8.9 &      &      &       &      \\
     & 0.85 &  42.3 &  5.9 &  77.7 &   7.9 &      &      &       &      \\
     & 0.95 &  37.7 &  6.2 &  44.8 &   6.2 &      &      &       &      \\
     & 1.05 &  54.4 &  6.6 &  36.0 &   5.4 &      &      &       &      \\
     & 1.15 &  36.0 &  5.6 &  28.2 &   4.8 &      &      &       &      \\
\cline{1-6}
\end{tabular}
\caption{Production cross sections for p-Au continued.}
\label{xs2au}
\end{table*}

\begin{table*}
\begin{tabular}{ccrrrrrrrr}
 & & \multicolumn{4}{c}{17.5 GeV/c} & \multicolumn{4}{c}{12.3 GeV/c} \\
$\cos\theta$ & $p$ 
& $d^2\sigma/dpd\Omega$ & error(stat) & $d^2\sigma/dpd\Omega$ & error(stat)
& $d^2\sigma/dpd\Omega$ & error(stat) & $d^2\sigma/dpd\Omega$ & error(stat)\\
& [GeV/c] & $\pi^+$ & \multicolumn{2}{c}{[mb/(GeV/c)]} & $\pi^-$
          & $\pi^+$ & \multicolumn{2}{c}{[mb/(GeV/c)]} & $\pi^-$ \\ \hline
0.95 & 0.12 & 101.4 &  7.3 & 188.1 &  9.9 & 154.4 & 11.0 & 202.2 & 12.6 \\
     & 0.17 & 201.7 & 31.4 & 252.1 & 31.7 & 229.8 & 43.4 & 206.9 & 38.4 \\
     & 0.25 & 362.3 &  8.7 & 428.2 &  9.4 & 361.2 & 10.6 & 380.5 & 10.9 \\
     & 0.35 & 499.2 & 10.2 & 527.5 & 10.5 & 473.4 & 12.2 & 495.6 & 12.5 \\
     & 0.45 & 582.7 & 11.0 & 573.6 & 10.9 & 517.9 & 12.7 & 502.9 & 12.5 \\
     & 0.55 & 608.5 & 11.3 & 608.5 & 11.3 & 589.4 & 13.6 & 442.8 & 11.8 \\
     & 0.65 & 638.6 & 11.5 & 604.2 & 11.2 & 600.5 & 13.7 & 496.9 & 12.5 \\
     & 0.75 & 596.4 & 11.1 & 564.5 & 10.8 & 570.3 & 13.4 & 466.5 & 12.1 \\
     & 0.85 & 611.1 & 11.3 & 549.5 & 10.7 & 555.1 & 13.2 & 441.7 & 11.8 \\
     & 0.95 & 576.4 & 11.0 & 524.0 & 10.4 & 514.9 & 12.7 & 382.0 & 10.9 \\
     & 1.05 & 610.5 & 11.3 & 471.2 &  9.9 & 477.5 & 12.2 & 370.7 & 10.8 \\
     & 1.15 & 523.7 & 10.4 & 443.5 &  9.6 & 457.3 & 12.0 & 350.9 & 10.5 \\
\hline
0.85 & 0.12 & 186.2 &  9.8 & 259.1 & 11.6 & 167.4 & 11.4 & 228.7 & 13.4 \\
     & 0.17 & 262.9 & 22.3 & 324.6 & 22.3 & 175.0 & 48.8 & 267.7 & 29.0 \\
     & 0.25 & 349.9 &  8.5 & 366.6 &  8.7 & 305.6 &  9.8 & 309.9 &  9.9 \\
     & 0.35 & 401.1 &  9.1 & 394.3 &  9.1 & 371.1 & 10.8 & 324.5 & 10.1 \\
     & 0.45 & 395.9 &  9.1 & 349.3 &  8.5 & 323.8 & 10.1 & 332.6 & 10.2 \\
     & 0.55 & 333.3 &  8.3 & 302.9 &  7.9 & 334.2 & 10.2 & 295.7 &  9.6 \\
     & 0.65 & 286.3 &  7.7 & 299.8 &  7.9 & 247.6 &  8.8 & 234.6 &  8.6 \\
     & 0.75 & 230.1 &  6.9 & 236.8 &  7.0 & 192.9 &  7.8 & 183.9 &  7.6 \\
     & 0.85 & 222.6 &  6.8 & 184.3 &  6.2 & 182.1 &  7.6 & 151.8 &  6.9 \\
     & 0.95 & 177.2 &  6.1 & 146.9 &  5.5 & 163.4 &  7.2 & 121.9 &  6.2 \\
     & 1.05 & 138.5 &  5.4 & 128.9 &  5.2 & 117.4 &  6.1 & 105.9 &  5.8 \\
     & 1.15 & 121.4 &  5.0 & 102.8 &  4.6 &  87.2 &  5.2 &  72.7 &  4.8 \\
\hline
0.75 & 0.15 & 229.7 & 16.4 & 260.9 & 16.6 & 246.4 & 17.5 & 141.5 & 54.9 \\
     & 0.25 & 317.0 &  8.1 & 324.3 &  8.2 & 294.2 &  9.6 & 258.6 &  9.0 \\
     & 0.35 & 298.3 &  7.9 & 282.2 &  7.7 & 274.2 &  9.3 & 228.5 &  8.5 \\
     & 0.45 & 254.3 &  7.3 & 237.3 &  7.0 & 243.7 &  8.7 & 206.9 &  8.0 \\
     & 0.55 & 195.4 &  6.4 & 184.5 &  6.2 & 174.2 &  7.4 & 178.8 &  7.5 \\
     & 0.65 & 156.5 &  5.7 & 154.6 &  5.7 & 126.8 &  6.3 & 123.8 &  6.2 \\
     & 0.75 & 123.2 &  5.1 & 119.3 &  5.0 &  92.5 &  5.4 & 105.8 &  5.8 \\
     & 0.85 &  89.2 &  4.3 &  79.5 &  4.1 &  84.8 &  5.2 &  63.5 &  4.5 \\
     & 0.95 &  71.8 &  3.9 &  73.9 &  3.9 &  64.9 &  4.5 &  57.2 &  4.2 \\
     & 1.05 &  36.0 &  2.7 &  43.0 &  3.0 &  52.9 &  4.0 &  37.3 &  3.4 \\
     & 1.15 &  50.5 &  3.2 &  47.0 &  3.1 &  39.2 &  3.5 &  30.8 &  3.1 \\
\hline
0.65 & 0.15 & 220.2 & 19.4 & 200.0 & 14.5 & 195.9 & 22.9 & 198.8 & 33.7 \\
     & 0.25 & 267.8 &  7.5 & 237.0 &  7.0 & 220.6 &  8.3 & 234.5 &  8.6 \\
     & 0.35 & 206.3 &  6.6 & 194.3 &  6.4 & 221.3 &  8.3 & 205.6 &  8.0 \\
     & 0.45 & 177.5 &  6.1 & 173.0 &  6.0 & 166.0 &  7.2 & 165.7 &  7.2 \\
     & 0.55 & 124.9 &  5.1 & 127.2 &  5.1 & 136.4 &  6.5 & 126.0 &  6.3 \\
     & 0.65 & 102.1 &  4.6 &  90.8 &  4.3 &  76.2 &  4.9 &  87.9 &  5.2 \\
     & 0.75 &  59.8 &  3.5 &  57.1 &  3.4 &  56.5 &  4.2 &  52.2 &  4.0 \\
     & 0.85 &  47.8 &  3.2 &  45.3 &  3.1 &  30.6 &  3.1 &  28.7 &  3.0 \\
     & 0.95 &  27.1 &  2.4 &  27.0 &  2.4 &  28.5 &  3.1 &  28.5 &  3.0 \\
     & 1.05 &  21.6 &  2.1 &  24.6 &  2.3 &  22.1 &  2.6 &  18.1 &  2.4 \\
     & 1.15 &  15.2 &  1.8 &  17.7 &  1.9 &   6.3 &  1.8 &   7.6 &  1.7 \\
\hline
\end{tabular}
\caption{Production cross sections for p-Cu}
\label{xscu}
\end{table*}

\begin{table*}
\begin{tabular}{ccrrrrrrrr}
 & & \multicolumn{4}{c}{17.5 GeV/c} & \multicolumn{4}{c}{12.3 GeV/c} \\
$\cos\theta$ & $p$
& $d^2\sigma/dpd\Omega$ & error(stat) & $d^2\sigma/dpd\Omega$ & error(stat)
& $d^2\sigma/dpd\Omega$ & error(stat) & $d^2\sigma/dpd\Omega$ & error(stat)\\
& [GeV/c] & $\pi^+$ & \multicolumn{2}{c}{[mb/(GeV/c)]} & $\pi^-$
          & $\pi^+$ & \multicolumn{2}{c}{[mb/(GeV/c)]} & $\pi^-$ \\ \hline
0.95 & 0.12 &  16.8 & 1.2 &  26.0 & 1.5 &       &     &       &     \\
     & 0.15 &       &     &       &     &  38.1 & 1.5 &  34.0 & 1.4 \\
     & 0.17 &  31.2 & 4.7 &  47.3 & 5.1 &       &     &       &     \\
     & 0.25 &  65.3 & 1.5 &  75.2 & 1.6 &  46.2 & 1.7 &  56.0 & 1.8 \\
     & 0.35 &  95.8 & 1.8 &  99.1 & 1.9 &  81.2 & 2.2 &  78.7 & 2.2 \\
     & 0.45 & 124.2 & 2.1 & 119.7 & 2.1 & 110.7 & 2.6 &  88.3 & 2.3 \\
     & 0.55 & 137.2 & 2.2 & 127.7 & 2.1 & 110.6 & 2.6 &  99.3 & 2.4 \\
     & 0.65 & 140.8 & 2.2 & 129.1 & 2.1 & 111.3 & 2.6 & 102.4 & 2.5 \\
     & 0.75 & 142.2 & 2.2 & 127.5 & 2.1 & 114.0 & 2.6 & 100.0 & 2.4 \\
     & 0.85 & 141.1 & 2.2 & 125.2 & 2.1 & 109.1 & 2.5 &  92.8 & 2.3 \\
     & 0.95 & 132.2 & 2.2 & 119.6 & 2.1 & 115.3 & 2.6 &  89.0 & 2.3 \\
     & 1.05 & 134.3 & 2.2 & 112.1 & 2.0 & 105.5 & 2.5 &  87.8 & 2.3 \\
     & 1.15 & 119.1 & 2.1 & 102.4 & 1.9 & 104.4 & 2.5 &  74.4 & 2.1 \\
\hline
0.85 & 0.15 & 36.3 & 1.1 & 37.9 & 1.2 & 32.2 & 1.4 & 32.9 & 1.4 \\
     & 0.25 & 54.5 & 1.4 & 58.1 & 1.4 & 53.3 & 1.8 & 47.1 & 1.7 \\
     & 0.35 & 72.3 & 1.6 & 69.2 & 1.6 & 66.1 & 2.0 & 58.2 & 1.9 \\
     & 0.45 & 77.6 & 1.7 & 72.5 & 1.6 & 63.9 & 1.9 & 62.0 & 1.9 \\
     & 0.55 & 66.6 & 1.5 & 71.8 & 1.6 & 55.3 & 1.8 & 50.5 & 1.7 \\
     & 0.65 & 60.3 & 1.5 & 55.9 & 1.4 & 50.0 & 1.7 & 47.4 & 1.7 \\
     & 0.75 & 50.2 & 1.3 & 44.4 & 1.3 & 37.9 & 1.5 & 33.1 & 1.4 \\
     & 0.85 & 40.1 & 1.2 & 35.8 & 1.1 & 39.7 & 1.5 & 29.4 & 1.3 \\
     & 0.95 & 33.1 & 1.1 & 31.2 & 1.0 & 27.1 & 1.3 & 25.9 & 1.2 \\
     & 1.05 & 24.0 & 0.9 & 26.0 & 1.0 & 23.3 & 1.2 & 16.9 & 1.0 \\
     & 1.15 & 23.3 & 0.9 & 19.4 & 0.8 & 16.7 & 1.0 & 15.7 & 1.0 \\
\hline
0.75 & 0.15 & 33.7 & 1.1 & 33.2 & 1.1 & 32.6 & 1.4 & 28.6 & 1.3 \\
     & 0.25 & 55.5 & 1.4 & 52.1 & 1.4 & 40.1 & 1.5 & 42.3 & 1.6 \\
     & 0.35 & 53.9 & 1.4 & 50.0 & 1.3 & 45.4 & 1.6 & 46.3 & 1.7 \\
     & 0.45 & 48.1 & 1.3 & 42.8 & 1.2 & 36.1 & 1.5 & 37.6 & 1.5 \\
     & 0.55 & 35.0 & 1.1 & 40.4 & 1.2 & 28.7 & 1.3 & 28.3 & 1.3 \\
     & 0.65 & 29.1 & 1.0 & 27.4 & 1.0 & 21.7 & 1.1 & 21.9 & 1.1 \\
     & 0.75 & 22.1 & 0.9 & 20.6 & 0.9 & 21.7 & 1.1 & 14.5 & 0.9 \\
     & 0.85 & 18.5 & 0.8 & 14.5 & 0.7 & 14.4 & 0.9 & 15.9 & 1.0 \\
     & 0.95 & 14.7 & 0.7 & 12.6 & 0.7 & 11.4 & 0.8 &  8.3 & 0.7 \\
     & 1.05 &  9.9 & 0.6 &  9.9 & 0.6 &  6.9 & 0.6 &  6.5 & 0.6 \\
     & 1.15 &  7.1 & 0.5 &  6.0 & 0.5 &  4.6 & 0.5 &  4.3 & 0.5 \\
\hline
0.65 & 0.15 & 31.3 & 1.1 & 29.8 & 1.0 & 23.6 & 1.2 & 28.1 & 1.3 \\
     & 0.25 & 45.8 & 1.3 & 40.9 & 1.2 & 29.9 & 1.3 & 36.4 & 1.5 \\
     & 0.35 & 34.5 & 1.1 & 38.2 & 1.2 & 30.4 & 1.3 & 29.7 & 1.3 \\
     & 0.45 & 33.4 & 1.1 & 31.0 & 1.0 & 24.2 & 1.2 & 21.3 & 1.1 \\
     & 0.55 & 24.2 & 0.9 & 22.0 & 0.9 & 19.0 & 1.1 & 17.5 & 1.0 \\
     & 0.65 & 17.7 & 0.8 & 16.7 & 0.8 & 10.4 & 0.8 & 12.3 & 0.9 \\
     & 0.75 & 12.1 & 0.7 & 10.7 & 0.6 & 10.5 & 0.8 &  8.6 & 0.7 \\
     & 0.85 &  9.0 & 0.6 &  8.2 & 0.5 &  5.4 & 0.6 &  6.3 & 0.6 \\
     & 0.95 &  5.1 & 0.4 &  5.8 & 0.5 &  7.4 & 0.7 &  4.8 & 0.5 \\
     & 1.05 &  2.7 & 0.3 &  3.8 & 0.4 &  2.7 & 0.4 &  3.0 & 0.4 \\
     & 1.15 &  1.8 & 0.3 &  2.5 & 0.3 &  2.7 & 0.4 &  1.7 & 0.3 \\
\hline
\end{tabular}
\caption{Production cross sections for p-Be}
\label{xsbe}
\end{table*}

\end{document}